# Single and Many Particle Correlation Functions and Uniform Phase Bases for Strongly Correlated Systems.

C W M Castleton† and M W Long.*

†Dept. Of Physics, University of Sheffield, Sheffield, S3 7RH, U.K.

*Dept. of Physics, University of Birmingham, Edgbaston, Birmingham, B15 2TT, U.K.

**Abstract.**

The need for suitable many or infinite fermion correlation functions to describe strongly correlated systems is discussed, and the question linked to the need for a correlated basis, in which the ground state *may* be positive definite for certain low dimensional geometries. In seeking a positive definite basis a particular trial basis is proposed, based on that for hard core bosons in pure one dimensional systems. Single particle correlations in this basis are evaluated for the case of the ground state of a quasi-1D Hubbard in the limit of extreme correlation. The model is a strip of the 2D square lattice wrapped around a cylinder, and is related to a ladder geometry with periodic boundary conditions along it's edges. This is done using both a novel mean field theory and exact diagonalisation, and the basis is indeed found to be well suited for examining (quasi)-order in the model. The model has a paramagnetic region and a Nagaoka ferromagnetic region. In the numerical calculation the correlation function in the paramagnetic phase has power law decay and the charge motion is qualitatively hard core bosonic. The mean field leads to an example of a BCS type model with single particle bosonic long range order.

## 1. Introduction.

There is currently a great deal of interest in the question of the movement of charge in low dimensional strongly correlated electron systems, partly inspired by the discovery in 1986 of the high temperature superconductors.[1] One of the difficulties in describing such systems is the invention and evaluation of suitable correlation functions. The problem is to decide what correlation functions most aptly measure the charge movement and order (if any) present in the ground state, and the elementary excitations.

A linked question is, what is the most appropriate basis to use in representing the ground state and measured correlation functions? This latter is a complex issue, as the most appropriate basis for describing the ground state and elementary excitations may not be the most appropriate for describing probes of the system. A clear case in point is that of photoemission from a spin-charge separated system. A spin-charge separated system is presumably best described using a basis involving separate operators for spin and charge. Photoemission, however, involves the ejection of a complete electron, so predicting the photoemission spectrum of a spin-charge separated system would involve taking convolutions over separate spin and charge wavefunctions, described using separate spin and charge operators. Clearly not only an appropriate basis is required, but also the transformations between it and the bases appropriate to experimental probes.

In seeking to develop an understanding of low dimensional correlated systems one *could* start by using one or two particle correlation functions, expressed in the most natural basis to work in - usually a simple fermionic one. The question is then whether or not this is sufficient to describe the behaviour of the system. This is clearly not always the case. A simple example is the movement of domain walls in models of magnetism. A simple one dimensional Ising model can contain domain walls, which in one dimension are single particle excitations of the model. However, creating a domain wall involves changing all the spins in the domain - potentially a large number. Moving the domain wall from some site $i$ to some other site $i+n$ involves flipping $n$ spins. Hence a description in the "natural" basis requires at least $n$ fermion correlation functions. The key to the problem is that the motion, viewed in that basis, is collective. However, a collective basis also exists, in which we create and annihilate domain walls, using some $b_i^\dagger$ and $b_i$. Expressed in *this* basis the domain wall motion ia a simple one particle affair, requiring only one particle correlation functions.

The implication is that in describing systems which involve strong correlations and hence possible collective motion, we may need to use $n$ or even $\infty$ particle correlation functions if we wish to use the "natural" basis. Alternatively, we can seek a collective basis, in which the correlations become single particle entities. The latter would seem the most appropriate choice, at least for describing the ground state charge motion and possible order.

1D systems are known to be spin-charge separated[2]. In the one dimensional Hubbard model, for example, the ground state is uniform phase and the excitations are phase excitations, which are indeed collective. In other words, the wavefunction contains no nodes, (is positive definite) and excitations can be described as the introduction of nodes. The most appropriate basis for describing the ground state charge motion in this system is clearly the uniform phase (or "positive definite") basis, otherwise known as hard core bosons. The basis is well known from the study of the XY-model. The transformation between hard core bose creation and annihilation operators and those for fermions (needed for photoemission etc.) is also known - the Jordan-Wigner transformation (see section 2). This is a transformation applicable in 1D which maps spinless fermions $(f_i, f_i^\dagger)$ onto hard core bosons $(b_i, b_i^\dagger)$ and vice versa. In other words, it exactly gives the positive definite basis in terms of the fermionic one. It is clear that, for example, $\left\langle b_i^\dagger b_j^\dagger b_{i+n} b_{j+n} \right\rangle > \left\langle f_i^\dagger f_j^\dagger f_{i+n} f_{j+n} \right\rangle$ since there will be no phase cancellation for the hard core bosons. In other words the positive definite, uniform phase basis is also the maximal correlation function basis. This may prove useful in seeking to identify the equivalent positive definite basis in quasi-1D and 2D systems. As it is the maximal correlation function basis, any ground state order present will be seen most clearly in this basis, and order seen in this basis need not be easily detectable in other bases.

The reason for this interest is that it has been suggested[3] that the 2D Hubbard model may similarly be spin-charge separated, with a positive definite charge wavefunction, and possibly collective charge motion. How this might come about is still unclear, since in a 2D fermionic system the nodes are unavoidable. However, one could imagine a spin-charge separated system in which all the nodes (the fermi statistics) are swallowed up into the spin wavefunction in some way. With a suitable difference in energy scales between the two subsystems this could leave an effective node-less charge wavefunction. (A similar situation will be examined in this paper.) The question of whether this actually occurs still remains contentious. Investigations usually revolve around the Hubbard model and it's strong coupling limit, the tJ model.

Analytic attempts to answer the question have not so far proved conclusive. Some renormalisation based calculations[4] find fermi liquid behaviour anywhere above pure 1D, but some other authors disagree.[5] Others have attempted to use a numerical approach. (See [6] for a detailed review of analytic and numerical work). Results from exact diagonalisation (ED) and quantum monte carlo (QMC) simulations yield complex and often tantalising phase diagrams, but each have severe limitations. There are indications of pairing correlations, usually associated with regions of phase separation at larger $J/t$. Anti-ferromagnetism occurs at half filling, and possibly Nagaoka ferromagnetism[7] near to half filling at small $J/t$[6,8−12]. However, the system sizes involved are very small. In ED the largest is usually 26 sites[8,9], while QMC simulations[10−12] can manage slightly larger systems, arounnd (12×12) but results are dogged by the sign problem[7] and cannot reach low temperatures. Hence, direct ED and QMC calculations on the 2D Hubbard and tJ models are still too restricted, and finite size effects make the boundary conditions far too important to draw solid conclusions. In particular, there are no clear indications that the pairing correlations lead to superconductivity in either the Hubbard or the tJ models.[6] Nor is it clear whether the excitation spectrum is essentially fermionic or spin charge separated. More promising, perhaps, is the recent 2D momentum space DMRG calculation by Xiang[13] which almost exactly matches the accuracy of ED calculations at smaller system sizes ($\leq 26$ sites) and can also produce results on system sizes of (to date) (12×12). Quantitatively, results are in keeping with QMC, but are now true variational bounds without sign problems. Further work along this line may be promising.

Another line which also seems promising is the study of doped tJ and Hubbard models on ladder geometries - see review in [14]. Whilst these have direct relevance to ladder materials, such as $SrCu_2O_3$ and $Sr_2Cu_3O_5$, they can also be viewed as finite width strips taken from the 2D square lattice. The finite width means that charges can move around one another, so exchange statistics start to have some importance, unlike in pure 1D. Hence one might seek to use these models to infer the behaviour of the 2D models, by extrapolating results from either single strips[15,16] or from a series of wider and wider strips.[14−16,] Reasonable results are availible for (at least) 2 leg,[17−19] 3 leg[15,20] and 4 leg[21,22] ladders. As one would expect from renormalisation group arguments, they all spin-charge separate, but the excitation spectrums vary massively : systems with odd numbers of legs form Luttinger liquids, with gapless excitation branches for both spin and charge. Even leg systems, on the other hand, are gapless only for charge excitations, and have been characturised[14,17,23] as Luther-Emery liquids. This large difference makes extrapolating the 2D behaviour almost impossible. The problem can, in a sense, be blamed on boundary conditions and finite size effects again : the edges of the strip constitute an open boundary condition running the length of what is in reality a very narrow system.

This last point is one that we will address in the current work. The model we will study (q.v.) can be viewed as a ladder geometry with *periodic* boundary conditions along the edges. As such it is hoped that it will prove more useful than the ladders in predicting the 2D behaviour.

In seeking an answer to the question of spin-charge separation in 2D one must also find a suitable basis in which to describe a 2D spin-charge separated system with a positive definite charge wavefunction. Currently, we do not know what this basis is (though, some mathematical progress will be made in a subsequent paper[24],) and so we will NOT provide any kind of an answer to that question. Instead, we will simply move *towards* a 2D description, by looking at a particular strongly correlated quasi-1D Hubbard model, which does spin-charge separate, introduced in section 3. First, in section 2, we introduce the basis and correlation functions we intend to use. We wish to seek a positive definite basis for our quasi-1D system, which we anticipate will be collective, and to examine single particle correlations defined in this basis. Unfortunately we do not know what such a basis should be, so instead we use a spin-charge separated basis introduced earlier[25]. We then apply the Jordan-Wigner transformation to the charge subsystem, and use the resulting basis as a start point. The ground state charge motion and order are examined in this basis using an exotic mean-field technique in section 4, and exact diagonalisation in section 5. In section 6 we conclude.

## 2. The Basis And Correlation Functions.

Before we can write down the correlation functions, we must define the basis. The normal start point for fermionic systems is the fermionic basis $f^\dagger_{i,\sigma}$, $f_{i,\sigma}$. We wish to restrict ourselves to the limit of infinite Hubbard repulsion ($U \longrightarrow \infty$) so we project out the doubly occupied states :-

$$g^\dagger_{i,\sigma} = f^\dagger_{i,\sigma} \left(1 - f^\dagger_{i,\bar{\sigma}} f_{i,\bar{\sigma}}\right) \tag{1a}$$

$$g_{i,\sigma} = \left(1 - f^\dagger_{i,\bar{\sigma}} f_{i,\bar{\sigma}}\right) f_{i,\sigma} \tag{1b}$$

The spin-charge separated operators[25] may be written as

$$g^\dagger_{i_1,\sigma_1} g^\dagger_{i_2,\sigma_2} .... g^\dagger_{i_n,\sigma_n} |0\rangle \longrightarrow c^\dagger_{i_1} c^\dagger_{i_2} .... c^\dagger_{i_n} |\sigma_1 \sigma_2 .... \sigma_n\rangle \tag{2}$$

where $c^\dagger_{i_1}$ and $c_{i_1}$ are creation/annihilation operators for a single spinless fermionic charge at site $i_1$. The phase is chosen to be positive for wavefunctions in which the fermionic operators occur in the order $c^\dagger_0$, $c^\dagger_1$, $c^\dagger_2$, ...., $c^\dagger_n$, ... The spin degrees of freedom are described by a compressed spin chain $|\sigma_1 \sigma_2 .... \sigma_n\rangle$ acted on by ring exchange operators $R_{\alpha_1,\alpha_n}$. Note that $\alpha_1 \in Z$ labels the position of spin $\sigma_1 = \pm\frac{1}{2}$ in the compressed spin chain, *not* the actual lattice site on which it occurs. $R_{\alpha_1,\alpha_n}$ cyclically permutes all the spins from $\alpha_1$ to $\alpha_n$. The ring exchange operators are related to the more common Heisenberg spin operators $\hat{S}_\alpha$ by the relation

$$\hat{R}_{\alpha_n,\alpha_1} = \prod_{\beta=\alpha_1}^{\alpha_{n-1}} \left[\frac{1}{2} + 2\hat{S}_{\beta+1}.\hat{S}_\beta\right] \tag{3}$$

Any electron wavefunction can of course be expressed in terms of these operators, be it fermi liquid, or spin-charge separated (Luttinger or Luther-Emery liquid) or otherwise. However the operators are clearly more suited to describing spin-charge separated situations, and are most easily applied in 1D or quasi-1D where an obvious linear site labeling exists. An exact mapping between these new operators and the conventional operators was given in reference [25]. The operation of this mapping can be shown by example. The state

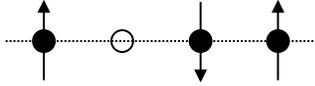

$$g^\dagger_{1,\uparrow} g^\dagger_{3,\downarrow} g^\dagger_{4,\uparrow} |0\rangle \qquad (4a)$$

in the standard notation, is now expressed as

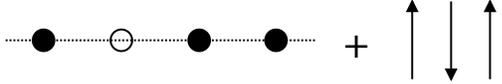

written as $\qquad c^\dagger_1 c^\dagger_3 c^\dagger_4 | \uparrow_1 \downarrow_2 \uparrow_3 \rangle. \qquad (4b)$

Similarly the operator

$$\sum_i g^\dagger_{i,\sigma} g_{i+2,\sigma} \quad \text{becomes} \quad \sum_{[i:\alpha]} c^\dagger_i c_{i+2} \left[ \left(1 - c^\dagger_{i+1} c_{i+1}\right) + c^\dagger_{i+1} c_{i+1} \hat{R}_{\alpha+1,\alpha} \right] \qquad (5)$$

The first term in square brackets deals with the situation in which there is no fermion at $(i+1)$ and hence the order along the compressed spin chain is unchanged. The second term deals with the situation in which $(i+1)$ is occupied so that two spins in the chain are cyclically permuted. The summations over $i$ (lattice sites) and $\alpha$ (positions in the compressed spin chain) are not, of course, independent and so care has to be taken in performing them.

This spin-charge separated basis is the one in which the model is most easily written (q.v.) and in which the mean field and numerical solutions will be most easily expressed. It is not, however, the positive definite basis we seek, in which we wish to describe the ground state charge motion. The charge degrees of freedom are here still expressed in terms of spinless fermion operators. Exactly what the required basis *is* for this model we do not know. As an approximation to it we will use the basis given by the Jordan-Wigner transformation.[26] This can be expressed as

$$b_i = c_i \exp\left( i\pi \sum_{j=0}^{i-1} c^\dagger_j c_j \right) \qquad (6)$$

where $b^\dagger_i$ and $b_i$ create and annihilate hard core bosons on the site $i$, and the phase is again chosen to be positive for wavefunctions created using fermionic operators in the order $c^\dagger_0$, $c^\dagger_1$, $c^\dagger_2$, ...., $c^\dagger_n$, ..... . The $i\pi$ phases from the exponents account for the difference in exchange statistics between fermions and bosons. Hence, for example,

$$b^\dagger_{i+2} b_i = c^\dagger_{i+2} c_i \exp\left( i\pi c^\dagger_{i+1} c_{i+1} \right)$$

$$= c^\dagger_{i+2} c_i \left( 1 - 2 c^\dagger_{i+1} c_{i+1} \right) \qquad (7)$$

This gives us a spin-charge separated basis in which we hope we have something close to a uniform phase representation for the charge wavefunction. It is approximation, however, because the Jordan-Wigner transformation works exactly only in srict 1D. The charge correlations we wish to measure are then single particle correlations in this basis -

$$B_n = \left\langle b_i^\dagger b_{i+n} \right\rangle \tag{8}$$

These may turn out to be not quite maximal, as this is *not* the pure 1D chain for which Jordan-Wigner is defined.

In terms of the original basis they are also collective. This can be seen by back transforming -

$$B_n = \left\langle c_i^\dagger c_{i+n} \exp\left(i\pi \sum_{j=i+1}^{i+n-1} c_j^\dagger c_j\right) \right\rangle$$

$$= \left\langle c_i^\dagger c_{i+n} \prod_{j=i+1}^{i+n-1} \left[1 - 2c_j^\dagger c_j\right] \right\rangle \tag{9}$$

Hence, written in the spinless fermionic basis this is a complex n-particle correlation function, describing the movement of a charge over n sites, subject to the positions of the charges in between. However, in the Jordan-Wigner basis it measures the mobility of what, if this was a pure 1D system, would be hard core bose quasi-particles. Alternatively it can be viewed as measuring the movement of a bare charge over n sites, irrespective of fermi exchange statistics. If $B_n$ is long ranged ($\lim_{n\to\infty} B_n \neq 0$) in a particular ground state it means that thereare long ranged charge fluctuations, but this does not neccasarily mean that the state is superconducting. This is because, firstly, we will present no information on the excitation spectrum of the model, so we can only *hope* that the low lying charge excitations are gapless, with the same statistics and mobility. Secondly, even if that is the case, we also require a macroscopic number of charges to be involved. However, if $\lim_{n\to\infty} B_n$ is zero it *does* mean that there is *no* order, and that the state is *not* superconducting.

If the correlation function in (8) is applied to a non-interacting free electron gas it is found to decay[27] with the power law $\sim \frac{1}{n^{\frac{1}{2}}}$.

## 3. The $t_1 - t_3$ Model.

We are interested in the 2D square lattice, in the extreme strongly coupled limit, $U = \infty$, $J = 0$. As mentioned above, a popular approach to the problem of the 2D square lattice is the study of (flat) ladder geometries. Some alternatives were introduced in reference [28], having different boundary conditions, and so providing alternative scalings to 2D. The one we study here is equivalent to wrapping a section of the 2D square lattice (with nearest neighbour bonds only) around a cylinder. It can be viewed as a multi-leg ladder with periodic (instead of open) boundary conditions parallel to the main axis. We also introduce a one lattice site displacement at this boundary. This results in a single spiral of sites running along the cylinder, (see figure 1,) with two distinct types of bonds, one around and one along. (These may be viewed as to some degree analagous to the perpendicular and parallel bond directions in the ladder geometries.)

We elect to label our sites along this spiral. This gives us a useful 1D labeling scheme for our quasi-1D model, in which each site is connected to two nearest neighbours, and to two $n^{th}$ nearest neighbours. It should be noted, however, that both of these arise from nearest neighbour bonds of the 2D square lattice (*not* next nearest neighbour bonds) and so, ideally, we would like to solve the models with the hopping parameters equal in these two directions.

What we have described is a homologous series of $t_1$-$t_n$ models, with $n$ odd. $n$ is proportional to the diameter of the cylinder, so the $n \longrightarrow \infty$ limit recovers the 2D lattice. Note that it is the bipartite nature of the 2D square lattice that forces $n$ to be odd. Consideration of longer range hopping in pure *1D* leads naturally to the consideration of the $t_1$-$t_2$ model. However, we are interested in 2D, not 1D, and $t_1$-$t_2$ actually corresponds to a strip from the 2D triangular lattice, where non-bipartite frustration leads to very different physics from that in which we are interested. Thus, the first model in the series is $t_1$-$t_3$, involving first and third nearest neighbour hopping terms at each site :-

$$\hat{H} = -t_1 \sum_{i,\sigma}[g^\dagger_{i,\sigma}g_{i+1,\sigma} + c.c.] - t_3 \sum_{i,\sigma}[g^\dagger_{i,\sigma}g_{i+3,\sigma} + c.c.] \qquad (10)$$

Since all the $t_1$-$t_n$ models are one dimensional in a renormalisation group sense they will all exactly spin-charge separate. Written in terms of our spin-charge separated operators $t_1$-$t_3$ becomes

$$\hat{H}_{t_1,t_3} = \begin{aligned} &-t_1 \sum_i \left[c^\dagger_i c_{i+1} + c.c.\right] \\ &-t_3 \sum_i \left[c^\dagger_i c_{i+3}(1 - c^\dagger_{i+1}c_{i+1})(1 - c^\dagger_{i+2}c_{i+2}) + c.c.\right] \\ &-t_3 \sum_{[i:\alpha]} \left[c^\dagger_i c_{i+3}\left(c^\dagger_{i+1}c_{i+1}(1 - c^\dagger_{i+2}c_{i+2}) + (1 - c^\dagger_{i+1}c_{i+1})c^\dagger_{i+2}c_{i+2}\right)\hat{R}_{\alpha+1,\alpha} + c.c.\right] \\ &-t_3 \sum_{[i:\alpha]} \left[c^\dagger_i c_{i+3}c^\dagger_{i+1}c_{i+1}c^\dagger_{i+2}c_{i+2}\hat{R}_{\alpha+2,\alpha} + c.c.\right] \end{aligned}$$
(11)

In terms of the spiral labeling the first term is simply nearest neighbour hopping and so involves no change in the compressed spin chain. The other terms are for third neighbour hopping over zero, one and two intermediate charges respectively. Hence the second term involves no ring exchange, whist the third and fourth terms involve ring exchanges of two and three spins respectively.

The $t_1$-$t_n$ series of models has been examined previously[25,28] when an exact perterbation theory in the limit $\frac{t_3}{t_1} \longrightarrow 0$ was presented. This calculation was for an infinite chain at T = 0K. For the $t_1$-$t_3$ model a paramagnetic state was found below a critical filling[28] $n_c = 0.6675045$, and a Nagaoka[7] ferromagnetic state above it, in keeping with some of the numerical results for the 2D square lattice[8], for example. It should be noted, however that in this instance (unlike in the numerical results) this is the Nagaoka state in the thermodynamic limit of a finte hole density. It was found, however[25], that the value of $n_c$ was dependant upon $n$, and that $n_c \to 1$ as $n \to \infty$. Ie the ferromagnetic phase disappeared in the limit of 2D square lattice connectivity, (albeit with the "wrong" matrix elements.) Unfortunately we are unable to prove this away from the $\frac{t_3}{t_1} \longrightarrow 0$ limit. Here we will present evidence that the phase diagram of $t_1$-$t_3$ at least is qualitatively unaltered away from this limit, and that the same phases are found even in the physically interesting case of $\frac{t_3}{t_1} = 1$.

## 4. Mean Field Treatment Of The $t_1 - t_3$ Model.

The basic principle is to study the spin degrees of freedom in the mean field of the charge degrees of freedom, and the charge degrees of freedom in the mean field of the spins. This amounts to assuming that the spin and charge are not only separated (as expected in 1D) but "independent." In this way local correlations between the two are lost, in favour of an effective spin hamiltonian and an effective charge hamiltonian. In the limit $\frac{t_3}{t_1} \longrightarrow 0$ the mean field recovers the exact perterbation theory results[28]. The charge hamiltonian was

$$
\hat{H}_{charge} = \begin{array}{l} -t_1 \sum_i \left[ c_i^\dagger c_{i+1} + c.c. \right] \\ \\ -t_3 \sum_i \left[ c_i^\dagger c_{i+3}(1 - c_{i+1}^\dagger c_{i+1})(1 - c_{i+2}^\dagger c_{i+2}) + c.c. \right] \\ \\ -R_1 t_3 \sum_i \left[ c_i^\dagger c_{i+3} \left( c_{i+1}^\dagger c_{i+1}(1 - c_{i+2}^\dagger c_{i+2}) + (1 - c_{i+1}^\dagger c_{i+1}) c_{i+2}^\dagger c_{i+2} \right) + c.c. \right] \\ \\ -R_2 t_3 \sum_i \left[ c_i^\dagger c_{i+3} c_{i+1}^\dagger c_{i+1} c_{i+2}^\dagger c_{i+2} + c.c. \right] \end{array}
$$

(12)

where

$$R_m = \langle \hat{R}_{\alpha+m,\alpha} \rangle \tag{13}$$

The spin Hamiltonian was

$$\hat{H}_{spin} = \left[ N_1^0 + N_2^0 + N_3^0 \right] \mathcal{N}_s + \left[ N_2^1 + N_3^1 \right] \sum_\alpha \left[ \hat{R}_{\alpha+1,\alpha} + c.c. \right] + N_3^2 \sum_\alpha \left[ \hat{R}_{\alpha+2,\alpha} + c.c \right] \tag{14}$$

$N_n^m$ is the expectation value of the operator which moves a charge from $i$ to $i+n$ passing $m$ occupied sites on the way. To express it in terms of the $c_i^\dagger$ operators, we first define the occupancy of a site $j$ as $\chi_j = 0$ or $1$. Then the set $\{\chi_{i+1}, \chi_{i+2}, ..., \chi_{i+j}, ..., \chi_{i+n-1}\}$ indicates which of the sites between $i$ and $i+n$ are occupied. Hence $\sum_{j=1}^{n-1} \chi_{i+j} = m$. The operator which checks whether or not a particular site $i+j$ is occupied in the wavefunction is then $\left[ \left(1 - c_{i+j}^\dagger c_{i+j}\right)^{1-\chi_j} \left(c_{i+j}^\dagger c_{i+j}\right)^{\chi_j} \right]$. Finally, in evaluating $N_n^m$ we need to sum over all possible permutations of the set $\{..,\chi_j,..\}$, ie we sum over all possible ways to have $m$ sites occupied between $i$ and $i+n$. Writing the sum over permutations as $\sum_{\wp(\{..,\chi_j,..\})}$ we obtain

$$N_n^m = -t_n \langle c_i^\dagger c_{i+n} \hat{N}_{i,n}^m \rangle \tag{15}$$

where

$$\hat{N}_{i,n}^m = \sum_{\wp(\{\chi_1, \chi_2, \ldots, \chi_j, \ldots, \chi_n\})} \prod_{j=1}^{n-1} \left[ \left(1 - c_{i+j}^\dagger c_{i+j}\right)^{1-\chi_j} \left(c_{i+j}^\dagger c_{i+j}\right)^{\chi_j} \right] \delta(m - \sum_{j=1}^{n-1} \chi_{i+j}) \tag{16}$$

The two sets of equations were examined self consistently as a function of the filling $n_e$ and the ration $\frac{t_3}{t_1}$. The spin hamiltonian is actually the well known $J_1/J_2$ nearest and next nearest neighbour Heisenberg model. The charge hamiltonian is a little more complex, and requires a further approximation. A Bogoliubov-Hartree-Fock mean field approximation was used, in which pairing correlations, $\langle c_i^\dagger c_{i+n}^\dagger \rangle$, were allowed. This led to a BCS type model with charge pairing, which was solved using a Bogoliubov transform[29]. A variational phase diagram was found by comparing the energy of the charges subject to various trial spin wavefunctions. The spin states used were the Néel state, the Heisenberg ground state, and a ferromagnet. For the low spin states the ground state always involved pairing. Initial results were published in reference [30]. The phase diagram obtained is shown here in figure 2. Qualitatively almost nothing alters as we move away from the $\frac{t_3}{t_1} \longrightarrow 0$ limit. It was previously noted that the drift of the phase transition towards lower fillings as $\frac{t_3}{t_1}$ increases is probably due, at least in part, to the fact that the spin-charge separation mean field is variational for the low spin states, but *exact* for the ferromagnet.

Here, in an extension to the previous paper, the results are presented for the correlation function $B_n$ introduced in equations (8),(9). Since this actually measures correlations in the 1D hard core bose (Jordan-Wigner transform) basis it is worth noting that the effective charge hamiltonian becomes the hamiltonian for hard core bosons (XY model) on the $t_1$-$t_3$ lattice if we set $R_1 = -1$ and $R_2 = +1$. For these values, $B_n$ measures maximal correlation functions by definition. This will be used for comparison shortly.

In the mean field, $B_n$ becomes[24,31]

$$B_n = 2^{n-1} \begin{vmatrix} \alpha_1 - \delta_1 & \alpha_0 - \frac{1}{2} & \alpha_1 + \delta_1 & \cdots & \alpha_{n-2} + \delta_{n-2} \\ \alpha_2 - \delta_2 & \alpha_1 - \delta_1 & \alpha_0 - \frac{1}{2} & \cdots & \alpha_{n-3} + \delta_{n-3} \\ \alpha_3 - \delta_3 & \alpha_2 - \delta_2 & \alpha_1 - \delta_1 & \cdots & \alpha_{n-4} + \delta_{n-4} \\ \vdots & \vdots & \vdots & \ddots & \vdots \\ \alpha_{n-1} - \delta_{n-1} & \alpha_{n-2} - \delta_{n-2} & \alpha_{n-3} - \delta_{n-3} & \cdots & \alpha_0 - \frac{1}{2} \\ \alpha_n & \alpha_{n-1} - \delta_{n-1} & \alpha_{n-2} - \delta_{n-2} & \cdots & \alpha_1 - \delta_1 \end{vmatrix} \begin{matrix} e \\ v \\ e \\ n \end{matrix} \tag{17}$$

Here

$$\alpha_n = \langle c_i^\dagger c_{i+n} \rangle \tag{18a}$$

and

$$\delta_n = \langle c_i^\dagger c_{i+n}^\dagger \rangle \tag{18b}$$

Both can be calculated easily once the gap equations have been solved self consistently and the charge ground state found. The $|..|_{even}$ indicates that only terms containing even powers of $\delta$'s (such as $\delta_1^2$ or $\delta_1^3 \delta_2$) are included in the determinant. This can easily be achieved by adding two determinants together in which the signs of the $\delta$'s are all reversed. Hence

$$B_n = 2^{n-2} \left[ \begin{vmatrix} \alpha_1 - \delta_1 & \alpha_0 - \frac{1}{2} & \cdots & \alpha_{n-2} + \delta_{n-2} \\ \alpha_2 - \delta_2 & \alpha_1 - \delta_1 & \cdots & \alpha_{n-3} + \delta_{n-3} \\ \vdots & \vdots & \ddots & \vdots \\ \alpha_n & \alpha_{n-1} - \delta_{n-1} & \cdots & \alpha_1 - \delta_1 \end{vmatrix} + \begin{vmatrix} \alpha_1 + \delta_1 & \alpha_0 - \frac{1}{2} & \cdots & \alpha_{n-2} - \delta_{n-2} \\ \alpha_2 + \delta_2 & \alpha_1 + \delta_1 & \cdots & \alpha_{n-3} - \delta_{n-3} \\ \vdots & \vdots & \ddots & \vdots \\ \alpha_n & \alpha_{n-1} + \delta_{n-1} & \cdots & \alpha_1 + \delta_1 \end{vmatrix} \right] \quad (19)$$

Figure 3 shows $B_n$ evaluated for the hard core bosons, as a function of $n$, for $\frac{t_3}{t_1} = 1$ and filling $n_e = 0.5$ (quarter filling). The correlations tend to a finite limit of $B_n = 0.2284785 \pm 0.0000001$ as $n \longrightarrow \infty$. Also shown on the same axes are the more conventional correlation functions $\delta_1 = \left\langle c_i^\dagger c_{i+1}^\dagger \right\rangle = 0.1889107 \pm 0.0000001$, and $C_n = < c_i^\dagger c_{i+1}^\dagger c_{i+n} c_{i+n+1} >$. The latter is given in mean field by

$$C_n = \delta_1^2 - \alpha_n^2 + \alpha_{n-1}\alpha_{n+1} \quad (20)$$

and tends to the value of $\delta_1^2$ as $n \longrightarrow \infty$. Being maximal for hard core bosons, $B_n$ is indeed the largest.

Figure 4 shows $\lim_{n \to \infty} B_n$ versus $n_e$ at $\frac{t_3}{t_1} = 1$ for hard core bosons and two paramagnetic variational states:- charges moving subject to the Heisenberg and Néel states along the compressed spin chain. It should be noted that for charges moving in the ferromagnet there is no scaling of $B_n$, which occillates wildly in sign. This is not suprising as the charges here are known to be spinless fermions. For charges moving subject to the Heisenberg ground state along the spin chain (the variational ground state for $\frac{t_3}{t_1} = 1.0$ and $n_e \leq 0.587$) there is again no scaling behaviour for $B_n$ above $n_e = 0.5939$, and the gap ($\delta_1$) closes very rapidly at this point. This behaviour will be easier to explain in the context of the numerical calculations, q.v. It is not a problem here, however, as it occurs just above the point at which this state ceases to be the variational ground state. (Unfortunately problems with the numerics make producing accurate data for $0.5639 < n_e < 0.5939$ impossible.)

In figure 5 the values of $\lim_{n \to \infty} B_n$, $\lim_{n \to \infty} C_n$ and $\delta_1$ are plotted together for the variational ground state (Heisenberg) at $\frac{t_3}{t_1} = 1.0$. For the Néel case they are similar. Non-zero values ($\forall n$) of $B_n$, $C_n$ and all pairing correlations $\delta_n$, are seen for both low spin states, indicating that, for most values of $n_e$, charge pairs exist in the ground state, and have long range order. The order is of course a result of the Bogoliubov-Hartree-Fock mean field making it an independent charge model. Hence, a "large" number of charges can move freely through the lattice in the mean field ground state. This does not make the model a superconductor as the model is one dimensional, and hence the Mermin-Wagner theorum forbids a macroscopic number of charges being involved. It should be noted though, that, potentially, only the dimensionality prevents macroscopic numbers being involved.

As anticipated, $\lim_{n\to\infty} B_n$ is always the largest of the correlation functions measured. This suggests either that $b_i^\dagger, b_i$ is indeed the maximal correlation function basis, (and hence the positive definite one,) or that it is at least closer to it than the conventional basis involving Cooper pairs $c_i^\dagger c_{i+n}^\dagger$.

Finally, it should also be noted that this demonstrates that long ranged *single* particle charge fluctuations can occur in an BCS type model. This raises the question of whether or not a single particle description (in a suitable basis) can be found for other BCS states, and if so, would it make a better description of the ground state than the standard ones[24]?

## 5. Exact Diagonalisation Studies Of The $t_1 - t_3$ Model.

The $t_1$-$t_3$ model has also been examined using exact diagonalisation of 12 and 16 site chains, for a variety of electron fillings, as a function of $\frac{t_3}{t_1}$. Periodic or anti-periodic boundary conditions were used, as appropriate to the electron filling. The exact diagonalisation was done using the Lanczos algorithm, on SUN SPARK 1, 2 and 10 workstations. The first results obtained were for the total spin of the system at each set of values considered. The results are given in figures 6a and 6b. The total spin corresponding to ferromagnetism ($S_{max}$) varies with the number of electrons involved, ($N_e$) and if there is an odd number of electrons then there is always a spinon left over even in the lowest possible spin configurations. This is awkward to plot, so the quantity actually plotted is

$$S_{scaled} = \begin{cases} \frac{S}{S^{max}} = \frac{2S}{N_e} & \text{for } N_e \text{ even} \\ \frac{S-\frac{1}{2}}{S^{max}-\frac{1}{2}} = \frac{2S-1}{N_e-1} & \text{for } N_e \text{ odd} \end{cases} \quad (21)$$

This gives a value ranging from 0 for a pure paramagnet to 1 for a ferromagnet.

The structure of these results is very similar to that of figure 2 (the mean field results) with a ferromagnetic region near to half filling, and a paramagnetic ground state at lower fillings. The main difference is that here the transition does not sink to lower fillings as $\frac{t_3}{t_1}$ grows. (The non-zero spin state around $n_e = 0.5$, $\frac{t_3}{t_1} = 1.0$ is perhaps a finite size effect, or possibly a commensurability effect.) This indicates that the increase in the size of the ferromagnetic region seen previously was indeed purely a results of the approximation.

Since the results for 12 and 16 site systems are qualitatively identical, and quantitatively almost identical, only the results for 16 sites will actually be shown from this point.

Following the spin state results, charge-only wavefunctions were extracted from the full electron wave-function obtained in the exact diagonalisation. For a finite sized system this can be done whether the system is spin-charge separated or not. Most importantly, it is possible to do it in such a way as to exactly obtain the positive definite basis for this charge system. The amplitude $A_{\{\chi_i\}}$ used for the $n$ charge configuration $\{\chi_i\}$ was the root sum of squares of the amplitudes $a_{\{\chi_i\},\{\sigma_j\}}$ for each of the states in the full fermionic wavefunction having that particular charge configuration. Ie,

$$A_{\{\chi_i\}}^2 \prod_i \left(c_i^\dagger\right)^{\chi_i} |0\rangle \iff \sum_{\sigma_1}..\sum_{\sigma_j}..\sum_{\sigma_n} a_{\{\chi_i\},\{\sigma_j\}}^2 \prod_i \left(g_{i,\sigma_j}^\dagger\right)^{\chi_i} |0\rangle \qquad (22)$$

($g_{i,\sigma_j}^\dagger$ as defined in equation (1).) The choice of phase is arbitrary, but always choosing positive phase gives the positive definite basis. It is possible to do this exactly as we have only a finite sized system. However, limitations on system size mean that at this point we can extract little useful scaling information for $B_n$ directly. Instead, we first find effective charge models and then scale those. This is done by overlapping the extracted charge wavefunctions with the ground state of an effective charge model, and varying the parameters of the effective model to obtain the optimal fit. The effective model used was the charge model that arose in the mean field calculation

$$\hat{H}_{eff} = \begin{aligned}&-t'_1 \sum_i \left[c_i^\dagger c_{i+1} + c.c.\right]\\ &-t'_3 \sum_i \left[c_i^\dagger c_{i+3}(1 - c_{i+1}^\dagger c_{i+1})(1 - c_{i+1}^\dagger c_{i+1}) + c.c.\right]\\ &-R_1 t'_3 \sum_{[i:\alpha]} \left[c_i^\dagger c_{i+3}\left(c_{i+1}^\dagger c_{i+1}(1 - c_{i+2}^\dagger c_{i+2}) + (1 - c_{i+1}^\dagger c_{i+1})c_{i+2}^\dagger c_{i+2}\right) + c.c.\right]\\ &-R_2 t'_3 \sum_{[i:\alpha]} \left[c_i^\dagger c_{i+3} c_{i+1}^\dagger c_{i+1} c_{i+2}^\dagger c_{i+2} + c.c.\right]\end{aligned}$$

(23)

Appropriate choice of the parameters can make this an interacting or non-interacting spinless fermion model, an interacting or non-interacting hard core boson model, (XY model,) or anything in between. This includes charges moving subject to a Néel antiferromagnet or to a Heisenberg ground state along the compressed spin chain. The effective model ground states were again found using the Lanczos algorithm. The size of the optimal overlap is indicated in figure 7. The overlaps are generally very large, but do become smaller - to at worst 97.3% - near the phase transition itself.

The effective models can be classified by their proximity to one of the four exactly known points, as shown in figure 8. The four points, labeled A, B, C and D in the figure are :-

|   |   |   |   |   |
|---|---|---|---|---|
|   | A | Spinless fermions, | $R_1 = +1$, | $R_2 = +1$, |
|   | B | Hard core bosons, | $R_1 = -1$, | $R_2 = +1$, |
|   | C | Charges in Néel state, | $R_1 = 0$, | $R_2 = 0$, |
| and | D | Charges in Heisenberg ground state, | $R_1 = 1 - 2\ln 2$ = -0.386294361 | $R_2 = 1 - 6\ln 2 + \frac{9}{4}\zeta(3)$ = -0.454255051 |

where $\zeta(n)$ is the Riemann zeta function. (The values for the Heisenberg case come from [32] and [33].) Anything away from points A and B can be interpreted either as interacting spinless fermions or as interacting hard core bosons.

Using this classification scheme a phase diagram can be built up for the behaviour of the charges in the ground state of both 12 and 16 site systems, as shown in figure 9. As an example, a few points are indicated on figure 9 and their positions on axes $R_2$ vs $R_1$ are shown on figure 10, which should be read together with figures 8 and 9. Note that those classified as Heisenberg like are in fact often much more efficient than the charges moving in the Heisenberg ground state itself. By increasing the strength of the short ranged correlations $R_1$ and $R_2$ from around 0.386 and 0.243 to $\sim 1.0$, (presumably at the expense of higher $R_m$ correlations not involved in the Hamiltonian) they achieve better energies. It should also be noted that the charge motion in the entire paramagnetic region is qualitatively that of hard core bosons.

Once the effective models had been obtained they were themselves exactly diagonalised. This could be done for a variety of larger system sizes as there are fewer degrees of freedom in the effective charge-only models. This meant that finite size scaling of the results could also be done. The quantity evaluated was again the correlation function introduced in section 2:

$$B_n = \left\langle b_i^\dagger b_{i+n} \right\rangle \tag{24}$$

This was done using the Jordan-Wigner transformation again. The effects of the periodic boundary conditions on this number complicate matters. For an infinite 1D system, $B_n$ is expected to have power law behaviour. For a finite sized system with periodic boundary conditions the correlations never completely decay, so the system is ordered, and the order propagates both ways round the ring. We expect something like

$$B_n = A \left( n^{-\alpha} + (M-n)^{-\alpha} \right) \tag{25}$$

where $M$ is the number of sites on the ring, $A$ is some constant and $\alpha$ some power. We could now fit $\alpha(M)$ using (25), then scale with $M$, but this would be error prone. Instead we note that selecting just $n = \frac{M}{2}$ ($M$ even) reduces (25) to

$$B_n = 2A \left( \frac{M}{2} \right)^{-\alpha} \tag{26}$$

This can be scaled with $M$ fairly simply. A plot of $\left( B_n n^{(\alpha-1)} \right)$ against $\frac{1}{n}$ is used. For the correct value of $\alpha$ a line of zero gradient is obtained, which will show up at larger $n$, irrespective of finite size effects at smaller $n$. Away from this value of $\alpha$ the curve diverges or goes to zero. A variety of trial values are used - see for example figure 11, for pure hard core bosons at quarter filling. A value of $\alpha = 0.225 \pm 0.025$ is obtained.

Table 1 shows the results of $\alpha$ for the linear chain ($t_3 = 0$), and (with $t_3 = 1.0$) hard core bosons and charges in the Néel or Heisenberg ground states. If $B_n$ is calculated directly for spinless fermions the sign fluctuates wildly, with no scaling, just as in the mean field. This again indicates that Jordan-Wigner does not give a positive definite wavefunction for spinless fermions. This does not mean such a basis doesn't exist of course, indeed for a finite sized system it must. If a suitable basis could be found, power law decay of a single particle correlation function in it could perhaps be expected, albeit with a large power.

Similarly, there is again no scaling behaviour for charges moving in a Heisenberg ground state for $n_e \tilde{>} 0.5$ and for $\frac{t_3}{t_1} \sim 1.0$. For the Heisenberg ground state, $R_1$ has the same sign as for hard core bosons, but $R_2$ has the opposite sign. The point at which the scaling behaviour is expected to cease is when the Jordan-Wigner basis used for $B_n$ ceases to be qualitatively similar to the positive definite basis sought, adn hence the wavefunction has positive and negative parts of the same order. This in turn is the point at which hopping over 2 other charges, with the phase of $R_2$, becomes more important than hopping over 1, with the phase of $R_1$. This should indeed occur at $n_e = 0.5$.

Figure 12 gives the power laws obtained from scaling the effective models themselves. Points for which effective models were scaled are marked by a cross. Power law scaling of $B_n$ is found everywhere except where there is ferromagnetism and hence spinless fermionic charges. For all other values of the parameters power laws are found, emphasising the essential unity of the whole paramagnetic phase. The existance of power laws suggests that power law behaviour for the charge correlations might also be expected in the full quasi-infinite system. This is in keeping with the Hartree-Fock mean field results in section 4, except that the spurious long ranged order does not occur in the exact diagonalisations.

In some areas of the $\frac{t_3}{t_1}$ versus $n_e$ phase diagram (figure 12) the exponent in the power law for the effective model is actually smaller than that expected for hard core bosons. This means that the charge correlations found in the ground state of the effective model, and hence most likely in the original fermionic model, are longer ranged than in the $t_1$-$t_3$ hard core boson model. The longest range correlations occur at the same place that the overlap is smallest. This is to be expected, since the effective models used involve only local charge interactions, and will therefore have the most trouble describing situations where the range of the charge correlations is greatest.

## 6. Conclusions.

In seeking a better understanding of uniform phase bases for strongly correlated systems we have introduced a certain spin-charge separated basis. Since we do not know how to represent positive definite wavefunctions in (1+$\delta$) D or above we have tried to represent the charge wavefunction using operators arrived at using the Jordan-Wigner transformation. This would produce the desired positive definite representation *if* the system were in pure 1D. We have evaluated the single particle correlations $B_n$ in this basis - collective correlation functions in the "natural" fermionic basis. We have done this for a particular quasi-1D Hubbard model, the $t_1$-$t_3$ model, using an exotic mean field theory, and exact diagonalisation studies.

The mean field theory produces a BCS type model, with two ground state phases. Close to half filling a Nagaoka ferromagnetic state was found. At lower fillings a paramagnetic state was found, in which, in the mean field, there were gaps to single particle fermionic excitations, and $\lim_{n\to\infty} B_n$ was finite, indicating the existance of order. It was also larger in magnitude than the $\left\langle c_i^\dagger c_{i+n}^\dagger \right\rangle$ type pairing correlation functions considered conventionally. The exact diagonalisation studies produce the same picture, but without the spurious long range order permitted by the mean field. The charge motion in the paramagnetic region was seen to be qualitaively that of hard core bosons, and $B_n$ was found to decay with (measured) power laws. The phase diagram is similar to those proposed for the 2D square lattice Hubbard and tJ model in refs [6,8].

The $t_1$-$t_3$ model thus appears to form a qualitatively hard core bose charge ground state, with long range collective charge fluctuations. This does not constitute a macroscopic superconductor, as the Mermin-Wagner theorum does not allow order in 1D, and we also know nothing from the current work about the low lying modes. As a result we cannot say whether to classify the $t_1$-$t_3$ model as a Luttinger or a Luther-Emery liquid. However we can say that the uniform phase charge wavefunction seems to arise as a result of the spin-charge separation : the nodes are placed in the spin wavefunction, resulting in strong short range spin correlations which effectively absorb the fermi statistics. The numerical results indicate that this is done extremely efficiently.

This basis is probably not actually maximal, (equivalent to actually *being* the positive definite/uniform phase basis). However, since the correlations measured in it are the largest we can find we conclude that it is at least the closest to maximal that we have availible. In the case of the numerical results we do come closer - we have the positive definite basis exactly - but we cannot evaluate the order in this basis directly, as we do not have a single (or few) particle formulation of it.

The $t_1$-$t_3$ model is a concrete example of a model in which strongly correlated charge motion requires many fermion correlation functions for it's description. Here we managed with an $n$ particle correlation function. In the general case an infinite particle correlation function might be required. However, we have also found that single particle correlations in a suitable positive definite basis provide a better description. This clearly has implications for proposed 2D spin-charge separated systems.

The $t_1$-$t_3$ charge dynamics also form a concrete example of a BCS type model for which single particle correlation functions can be used in describing the ground state, as opposed to the two particle description normally used. This raises the question of whether or not a single particle order parameter could be used to describe other superconducting states. Half an answer is given - if $B_n$ is non-zero there *may* be a superconducting ground state (subject to the involvement of macroscopic numbers etc.) On the other hand, if $B_n = 0$ there will *not* be order and superconductivity.

None of our results tell us anything about the groundstate or excitation spectrum of the 2D square lattice Hubbard, tJ or even t models. However, it is notable that the mean field and exact diagonalisation phase diagrams for $t_1$-$t_3$ away from $\frac{t_3}{t_1} = 0$ are qualitatively and almost quantitatively the same as the exact perturbation theory results at $\frac{t_3}{t_1} \to 0$. In the $n \to \infty$ limit of the perterbation theory, (corresponding to 2D connectivity,) only the paramagnetic phase remained. One could suggest from this that the 2D square lattice t model should have nothing but paramagnetism anywhere away from half filling, with qualitatively hard core bosonic charge correlations, and maybe ordering. This would certainly be in keeping with much of the other work on the square lattice and ladder geometries, but this is pure conjecture. To proceed beyond conjecture we must apply these ideas above 1D. The problem with this is one of formulation. Here, power law order in 1D was seen (suggesting the possiblity of full long range order in >2D or perhaps ≥2D) because something close to the transformation to a positive definite basis is known (the Jordan-Wigner transformation). If suitable transformations could be found for other systems then perhaps full single particle long range order could again be seen. This would obviously facilitate the development of a new way to visualise and describe superconductivity in general, and would have wide-ranging application to the study of high temperature superconductivity in particular.[24]

## 6. References.

**Table 1.**
Table showing powers $\alpha$ for scaling of $B_n$ for the specified charge models. Results given $\pm 0.025$.

|  | $\alpha$ **For Power Laws At Given Fillings.** | | | | | |
|---|---|---|---|---|---|---|
| **Model.** | $n_e = \frac{1}{4}$ | $n_e = \frac{1}{3}$ | $n_e = \frac{1}{2}$ | $n_e = \frac{5}{8}$ | $n_e = \frac{2}{3}$ | $n_e = \frac{3}{4}$ |
| Linear Chain. | 0.500 | 0.500 | 0.500 | 0.500 | 0.500 | 0.500 |
| Hard Core Bosons. | 0.225 | 0.225 | 0.225 | 0.200 | 0.225 | 0.225 |
| Heisenberg. | 0.300 | 0.300 | 0.300 | - | - | - |
| Néel. | 0.325 | 0.275 | 0.250 | 0.300 | 0.375 | 0.475 |

**Figure 1.**
The first three $t_1$-$t_n$ models. ($n = 3, 5, 7$)

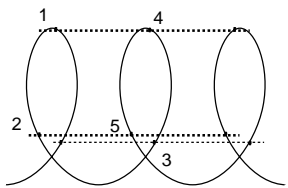

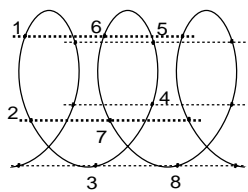

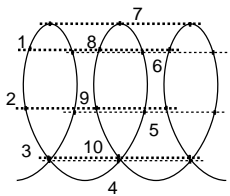

**Figure 2**
$\frac{t_3}{t_1}$ vs. n$_e$ phase diagram for the $t_1$-$t_3$ model from the mean field theory. From ref [30].

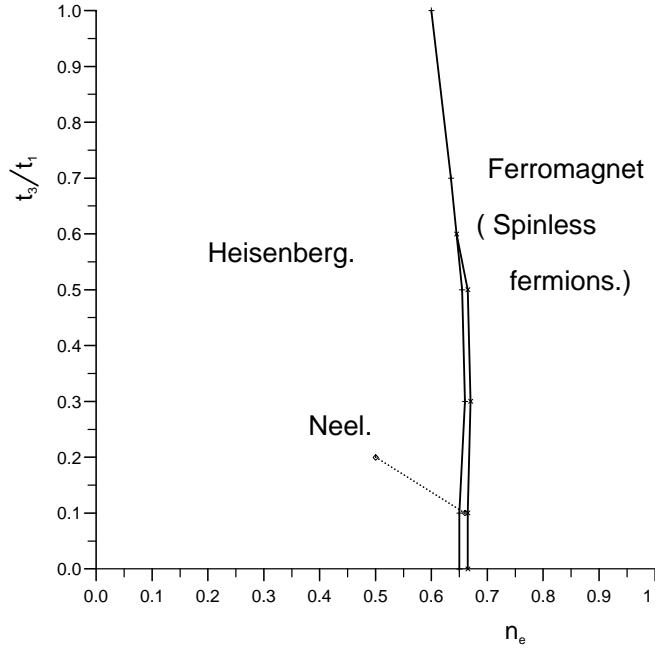

**Figure 3**
Scaling correlation functions with $n$, for hard core bosons. Filling n$_e$ = 0.5 and $\frac{t_3}{t_1} = 1$.
A: $B_n = \left\langle b_i^\dagger b_{i+n} \right\rangle$, B: $\delta_1 = \left\langle c_i^\dagger c_{i+1}^\dagger \right\rangle$, C: $\delta_1^2$, D: $< c_i^\dagger c_{i+1}^\dagger c_{i+n} c_{i+n+1} >$.

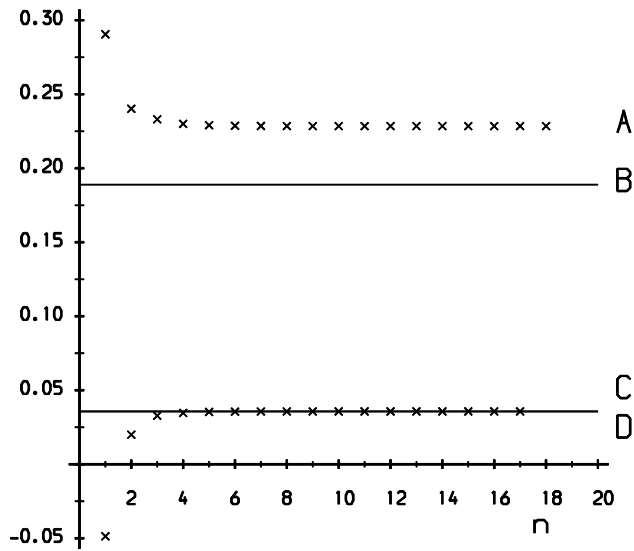

**Figure 4**
$\underset{n \longrightarrow \infty}{lim} B_n$ as a function of filling $n_e$ at $\frac{t_3}{t_1} = 1$.
A: Hard Core Bosons and charges in B: the Néel state and C: the Heisenberg ground state.

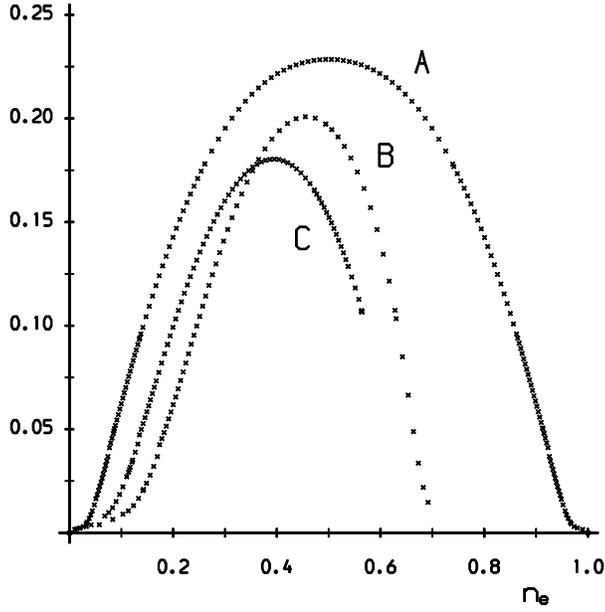

**Figure 5**
Correlation functions vs. $n_e$ for mean field variational ground state. (Charges in the Heisenberg ground state.)
A: $\underset{n \longrightarrow \infty}{lim} B_n$, B: $\delta_1 = \left\langle c_i^\dagger c_{i+1}^\dagger \right\rangle$, C: $\underset{n \longrightarrow \infty}{lim} < c_i^\dagger c_{i+1}^\dagger c_{i+n} c_{i+n+1} >$.

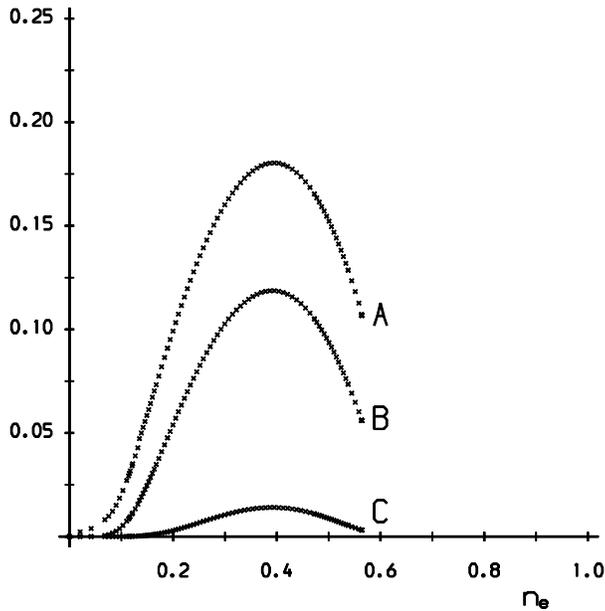

**Figure 6a**
$S_{scaled}$ spin phase diagram from exact diagonalisation of the $t_1$-$t_3$ model on 12 sites.

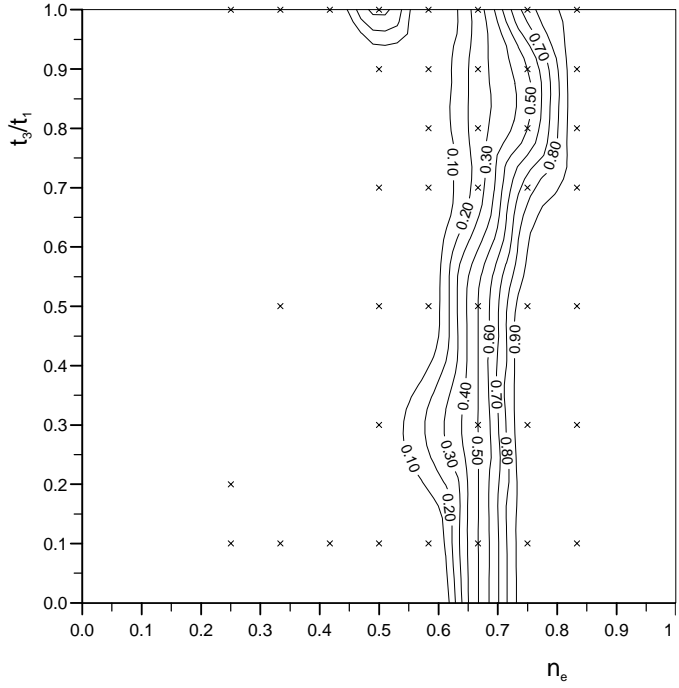

**Figure 6b**
$S_{scaled}$ spin phase diagram from exact diagonalisation of the $t_1$-$t_3$ model on 16 sites.

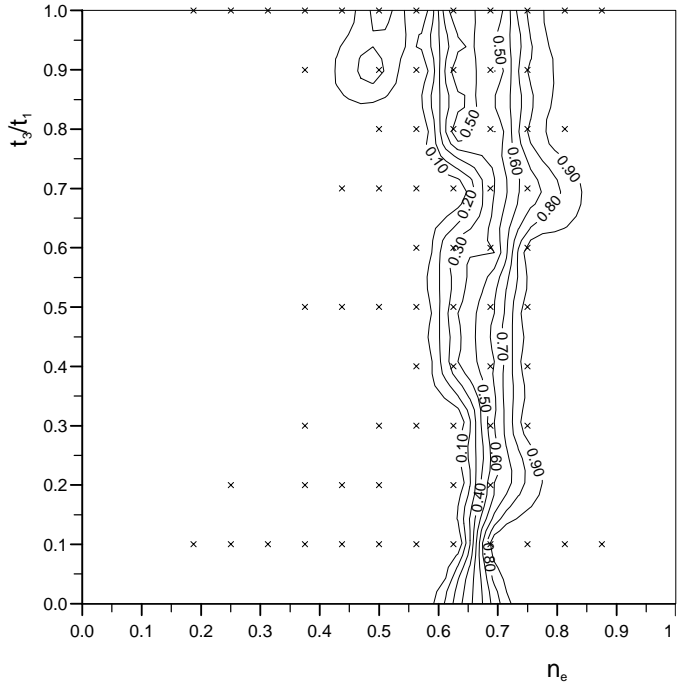

**Figure 7**
Error in effective model: (1 - the overlap) between the charges in the full fermionic ground state and in the effective charge model, on 16 sites.

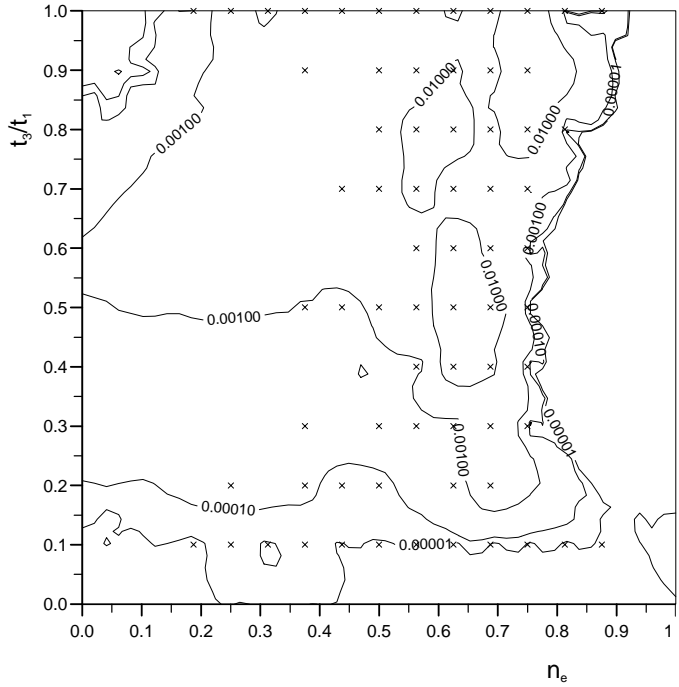

**Figure 8**
Classifications used for the effective charge models with $R_1$ and $R_2$.
"×" refers to an exactly known point - see main text.

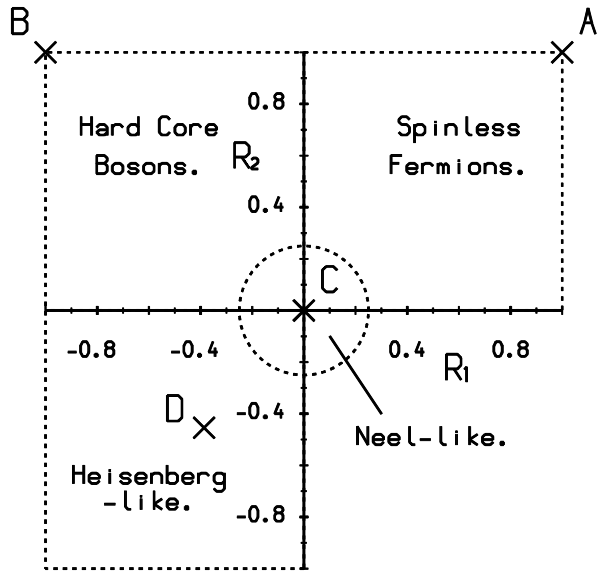

**Figure 9**
Phases in the fermionic model as characterised by the effective charge model.
The selected sample of points corresponds to those shown in figure 10, where their effective $R_1$ and $R_2$ values are plotted.

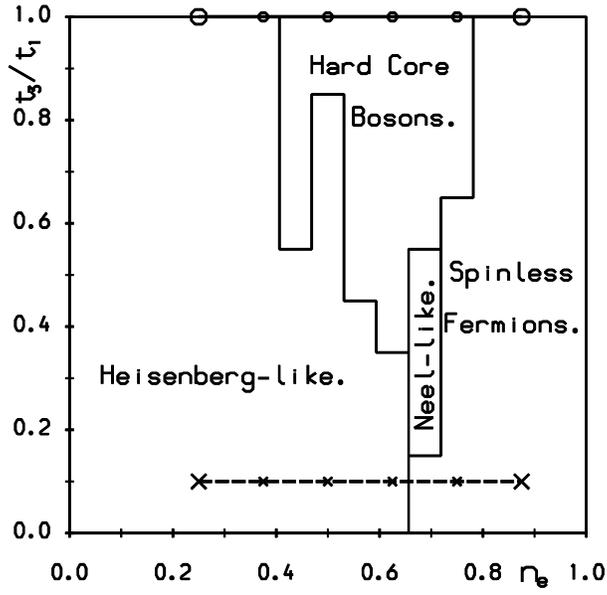

**Figure 10**
Effective charge model classifications of the points selected in figure 9. The points marked by circles lie along $\frac{t_3}{t_1} = 1.0$ in figure 9, and those with crosses lie along $\frac{t_3}{t_1} = 0.1$. The $n_e$ values are labeled.

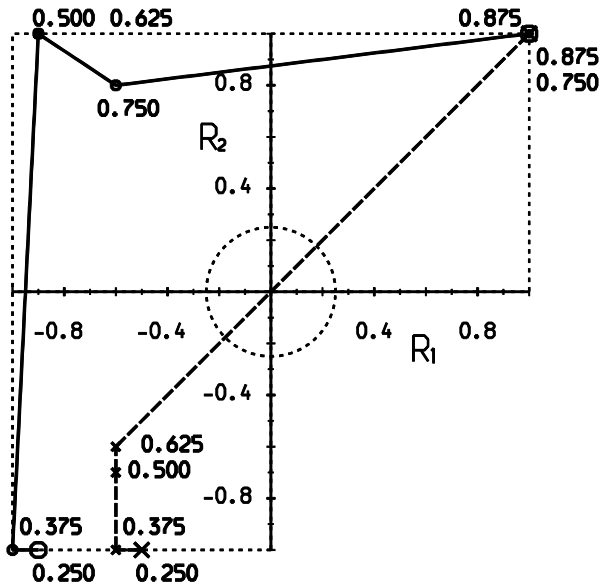

**Figure 11**
Scaling to find the power $\alpha$ for pure hard core bosons at $n_e = 0.5$, for the $t_1$-$t_3$ model.

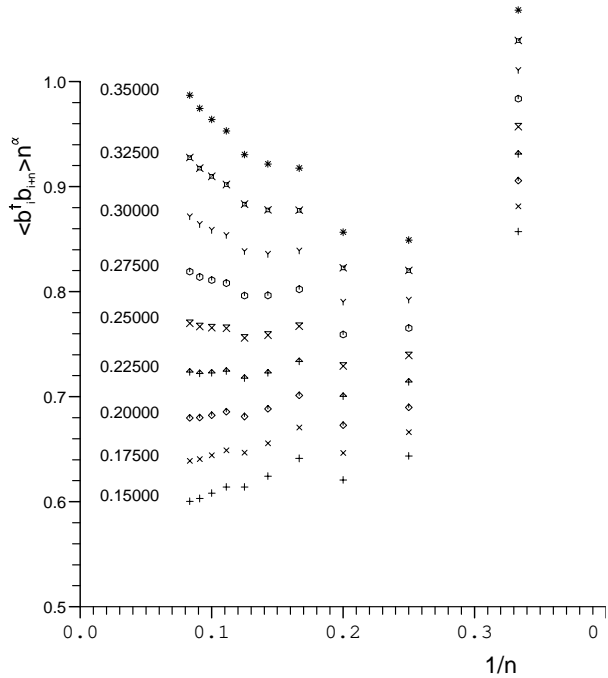

**Figure 12**
Powers $\alpha$ for the decay of $B_n$, from scaling the effective model. Plotted against $n_e$ and the ratio $\frac{t_3}{t_1}$ for the original model. Results are $\pm 0.025$.
The crosses mark the points at which $\alpha$ was evaluated. Note that the exact results[7] gave $\alpha = 0.5$ everywhere for $\frac{t_3}{t_1} = 0.0$. Near $n_e = 1.0$ the ground state of the model becomes fermionic and hence $B_n$ does not scale.

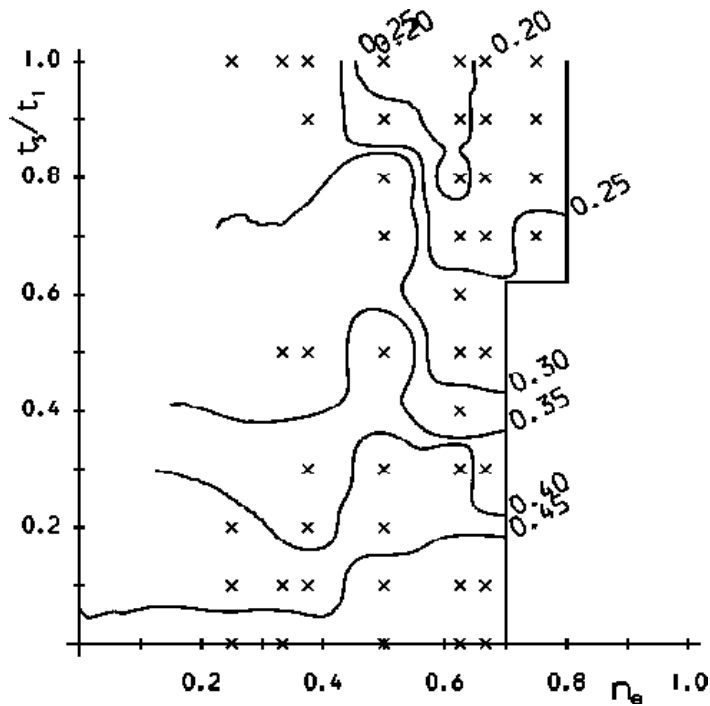